\documentstyle[epsf]{article}
\begin{document}

\begin{flushright}
\large{
                    KEK CP-034\\
                    KEK Preprint 95-114
}
\end{flushright}
\renewcommand{\baselinestretch}{1.5}
\large

\begin{center}
{\Large {\bf Improved QEDPS for Radiative Corrections\\
                 in $e^+e^-$ Annihilation} }
\vskip 0.8in
   T. Munehisa(*), J. Fujimoto, Y. Kurihara and Y. Shimizu
\vskip 0.3in

(*) Faculty of Engineering, Yamanashi University
\vskip 0.1in
Takeda, Kofu, Yamanashi 400, Japan
\vskip 0.2in
National Laboratory for High Energy Physics(KEK)
\vskip 0.1in
Oho 1-1 Tsukuba, Ibaraki 305, Japan
\end{center}

\vskip 0.5in
\begin{center}
{\bf ABSTRACT}
\end{center}
The generator QEDPS, a parton shower model in QED, developed by
the present authors has been improved. By a careful study on the shower
algorithm it was found that some finite contributions have been missed
in the original version. They are small, but cannot be neglected when 
the contribution of the soft photons is dominant. A method is presented
to correct these finite pieces and then a new generator, improved 
QEDPS, is proposed, which is able to make more precise prediction for
the processes observed in $e^+e^-$ annihilation as far as the initial
radiation is concerned.

\eject

\def\GeV{\hbox{GeV}}
\def\vec#1{{\bf{#1}}}

\noindent
{\bf Section 1. Introduction }

In detailed studies of  radiative corrections to $e^+ e^-$ 
processes one needs  Monte Carlo generators. Several authors have
proposed such generators so far\cite{gen}$-$\cite{ps2}.
We have already published QEDPS for radiative corrections in the 
$e^+ e^-$ annihilation\cite{ps1} and also in the Bhabha 
scattering\cite{ps2}. These generators are a QED version of the parton 
shower which has been originally developed for the study of 
QCD\cite{ll}. The algorithm is based on the $Q^2$-evolution according 
to the renormalization group equations, which enable us to sum up 
all the collinear singularities. 

The current version of QEDPS is written in the leading-logarithm(LL) 
approximation. Recent experiments at $e^+e^-$ colliders, however,
can provide very precise data and this in turn demands that the
theoretical uncertainty in a generator should be less than 1\% level.
In order to respond to this requirement we have been studying 
the complete next-to-leading-logarithm of the QEDPS like QCD\cite{nll}.
During this study we found that some finite contributions 
had been neglected in the original version. This takes place 
in the algorithm we adopted. 
These contributions are small, but actually cannot be ignored for 
the observables for which soft photons are important.
   
We explain how these contributions arises and how to do with them
in the next section. Their origin is directly connected to the
kinematics used in the generation algorithm. First we take the single 
cascade scheme which means that the electron with the momentum parallel 
to the axial gauge vector does not make showers. In section 3 the same 
problem will be discussed for the double cascade scheme. In this case 
any electron is able to make showers. Then one possible way to correct 
them is proposed and some results by the improved generator will be 
presented in section 4. Final section is devoted to summary and 
discussion.

\vskip 0.5in
\noindent
{\bf Section 2. $Q^2$-evolution }

The algorithm of the parton showers is based on the Altarelli-Parisi(AP)
equation which governs the $Q^2$-evolution of the structure function of
quarks and gluons in QCD. In QED the strong coupling $\alpha_s$ is 
replaced by that of QED $\alpha$ and the color factors are removed from
the AP equations. The equation for non-singlet $Q^2$-evolution 
is given by
\begin{equation}
  {d D(x,Q^2) \over d \ln Q^2} = {\alpha\over 2\pi}
\int_x^1 {dy \over y} P_+(x/y) D(y,Q^2) .
  \label{eq:AP2}
\end{equation}
\noindent
Here $D(x,Q^2) $ is the structure function of the electrons with $x$ 
being the electron momentum fraction and $P_+(x)=((1+x^2)/(1-x))_+$ is 
the split function. In the following we shall neglect the running 
effect of $\alpha$ only for the sake of simplicity. By taking the 
moments of this equation, one can easily obtain the solution 
\begin{equation}
   D(n,Q^2)=\int^1_0 dx x^{n-1} D(x,Q^2)  
         =A_n \exp\left( -{\alpha\over 2\pi } \gamma_n 
 \log(Q^2/Q^2_0)\right), 
 \label{eq:AP3}
\end{equation}
where $A_n$ is the integration constant and 
\begin{equation}
    \gamma_n=\int^1_0 dx x^{n-1}P_+(x) =
 { 3\over 2}-{1\over n}-{1\over (n+1)} -2\sum_{j=1}^{n-1} {1\over j }.  
  \label{eq:mom1}
\end{equation}
is the anomalous dimension.
Thus the $Q^2$-evolution of the structure function is determined by
\begin{equation}
   D(n,Q^2_1)/ D(n,Q^2_2)
    = \exp\left( {\alpha \over 2\pi}\gamma_n \log(Q^2_1/Q^2_2)\right).
  \label{eq:mom2}
\end{equation}

 Let us show the algorithm how to generate the parton showers.
First we have to regard Eq.(\ref{eq:AP2}) as an equation which
describes the stochastic process of emitting photons. For this
we modify the split function as follows\cite{ps1};
\begin{equation}
P_+(x)=\theta(1-\epsilon-x)P(x)
             -\delta(1-x)\int\nolimits_0^{1-\epsilon}dyP(x),\quad
                             P(x)={ 1+x^2 \over 1-x }. 
\end{equation}
Here $\epsilon$ is a small quantity and $\theta$ is the step function. 
Then the evolution equation can be cast into the form
\begin{eqnarray}
D(x,Q^2)&=& \Pi(Q^2,Q_s^2)D(x,Q_s^2)\nonumber\\
&+&{\alpha\over2\pi}\int\nolimits_{Q_s^2}^{Q^2}{dK^2\over K^2}
    \Pi(Q^2,K^2)\int\nolimits_x^{1-\epsilon}{dy\over y}
           P(y)D(x/y,K^2), \label{eq:intform}
\end{eqnarray}
where $Q_s^2$ is the minimum of $Q^2$ and $\Pi(Q^2,{Q'}^2)$ is 
the non-emission probability defined by
 \begin{equation}
 \Pi(Q^2,{Q'}^2) = \exp\left(- {\alpha\over 2 \pi} \int_{{Q'}^2}^{Q^2}
 { d K^2 \over K^2}
             \int_0^{1-\epsilon} d x P(x) \right).
  \label{eq:non}
\end{equation}

\noindent
The evolution equation in the integral form Eq.(\ref{eq:intform}) allows 
one to take it as that for stochastic process and the algorithm of the 
photon shower consists of the following steps.

(1) Set  $x_b=1 $, where $x_b$ becomes the fraction of the light-cone
    momentum of the electron after the end of the shower or just 
    before the annihilation.

(2) If a given random number $\eta $ is smaller than $\Pi(Q^2,Q_s^2)$,
    the evolution stops. If not, one finds the virtuality $K^2$ that 
    satisfies $\eta =\Pi(K^2,Q_s^2)$ with which a branching is made.

(3) Fix $x$ according to the probability $P(x)$ between 0 and 
    $1-\epsilon$. Then $x_b$ is replaced by $x_b x$.
    One should go to (2) by substituting $K^2$ into $Q_s^2$ and 
    repeat until it stops.

According to this algorithm the probability that no photons is emitted
during the evolution from $Q_s^2$ to $Q^2$ is $\Pi(Q^2,Q_s^2)$.
The probability for the single photon emission is
\begin{eqnarray}
 &&\int_{Q_s^2}^{Q^2} d K^2 \Pi(K^2,Q_s^2)  {\alpha\over 2 \pi}  
{1\over K^2}
             \int_0^{1-\epsilon} d x P(x)  \Pi(Q^2,K^2)  \nonumber \\
 &=& \Pi(Q^2,Q_s^2)  \int_{Q_s^2}^{Q^2} d K^2   {\alpha\over 2 \pi}
             {1\over K^2} 
             \int_0^{1-\epsilon} d x P(x), 
\end{eqnarray}
and for the double photons it is
\begin{eqnarray}
&& \int_{Q_s}^{Q^2} d K_2^2 \int_{Q_s^2}^{K_2^2} d K_1^2 
\Pi(K_1^2,Q_s^2)  {\alpha\over 2 \pi} {1\over K_1^2} 
             \int_0^{1-\epsilon} d x_1 P(x_1)  \Pi(K_2^2,K_1^2) 
\nonumber \\
   &&\times{\alpha\over 2 \pi}  {1\over K_2^2}
             \int_0^{1-\epsilon} d x_2 P(x_2)  \Pi(Q^2,K_2^2) 
\nonumber \\ 
&=&\Pi(Q^2,Q_s^2)  \int_{Q_s^2}^{Q^2} d K_2^2   {\alpha\over 2 \pi} 
 {1\over K_2^2}
             \int_0^{1-\epsilon} d x_2 P(x_2)   
  \int_{Q_s^2}^{K_2^2} d K_1^2   {\alpha\over 2 \pi}  {1\over K_1^2}
             \int_0^{1-\epsilon} d x_1 P(x_1)    . \nonumber \\
&&
\end{eqnarray}
\noindent
For the emission of any number of photons, we have similar 
expressions.
  
When one wants to see the distributions on $x_b$, the $\delta$-functions
are inserted into the integrands,
\begin{eqnarray}
   \delta(x_b - x_1 x_2...x_N). 
\end{eqnarray}
If one takes the moments, one gets a compact expression.
\begin{eqnarray}
    D(n,Q^2) = \exp\left( {\alpha\over 2 \pi} \int_{Q_s^2}^{Q^2}
{ d K^2\over K^2}\int_0^{1-\epsilon} d x P(x)(x^{n-1}-1) \right) ,
  \label{eq:mom3}
\end{eqnarray}
by noting  that 
\begin{eqnarray}
&&\int_0^1 dx_b x_b^{n-1}  \delta(x_b-x_1x_2\cdots x_N) =
 (x_1x_2 \cdots x_N)^{n-1}  ,\\
&& \int_{Q_s^2}^{Q^2} d K_N^2 \int_{Q_s^2}^{K_{N}^2} d K_{N-1}^2
       \cdots   
  \int_{Q_s^2}^{K_3^2} d K_2^2 \int_{Q_s^2}^{K_{2}^2} d K_{1}^2 
 = {1\over N!}\prod_i^N \int_{Q_s^2}^{Q^2} d K_i^2 . 
\end{eqnarray}

If we assume a very small constant value $\epsilon_0$ for the parameter
$\epsilon$, we reproduce Eqs.(\ref{eq:mom2}) because
the corresponding structure function becomes
\begin{eqnarray}
    D_0(n,Q^2)&=&\exp\left( {\alpha\over 2 \pi} \log(Q^2/Q^2_s)
  \int_0^{1-\epsilon_0} d x P(x)(x^{n-1}-1) \right) \\
   &=&\exp\left( {\alpha\over 2 \pi} \log(Q^2/Q^2_s)
             \gamma_n + O(\epsilon_0) \right).
  \label{eq:d0}
\end{eqnarray}
\noindent
No problem arises, if $\epsilon_0$ is a constant. In the parton shower 
algorithm to generate events, however, we have $K^2$-dependent 
$\epsilon$(see ref.\cite{ps1}). In QEDPS it is given by
\begin{eqnarray}
     \epsilon_K = Q_0^2/K^2 , 
\end{eqnarray}
which comes from the fact that the kinematical boundary restricts 
the fraction $x$ of the light-cone variable. Here $Q_0^2$ is a cutoff,
a fictitious mass of the emitted photon introduced to regulate the
infrared divergence.
  
In order to see whether the Monte Carlo simulation with the above
mentioned algorithm can reproduce the moment given by an analytic
expression Eq.(\ref{eq:d0}) we make a numerical comparison. We take 
$Q_0^2=Q^2_s=10^{-6} \GeV^2, Q^2=10^2 \GeV^2$ and $\alpha=1/20$. 
The last value is assumed only to magnify the finite contributions.
Figure 1 shows the both results from the simulation(points with
diamond) together with those from the analytic formula(solid curve).
Apparently there is a sizeable discrepancy. However this does not
imply that the cutoff $\epsilon_K$ is wrong.

This can be understood by looking at the ratio of the moments 
$D(n,Q^2)/D(n,Q_1^2)$. This agrees completely with the analytic 
results. Hence one concludes that the structure function with the cutoff
$\epsilon_K=Q_0^2/K^2 $ should give some finite contributions.
As long as the $Q^2$-evolution(the ratio at two different $Q^2$'s) is 
concerned, everything is correct. However if one is interested in the 
absolute value of the structure function, the algorithm is not 
sufficiently accurate and should be improved. The generated $D(n,Q^2)$
must be able to reproduce the results of perturbative calculation which
have been known in the literature as mentioned in Section 4. Thus one
can see that in the actual applications the finite contribution does 
not cause any problem in QCD, but in QED more careful study is required.

Let us now evaluate the finite contribution. First we define a structure
function $D_s(n,Q^2)$ with the cutoff $\epsilon_K$ as
\begin{eqnarray}
 D_s(n,Q^2)\equiv\exp\left( {\alpha\over 2 \pi} \int_{Q_0^2}^{Q^2} 
                     {d K^2\over K^2}
             \int_0^{1-\epsilon_K} d x P(x)(x^{n-1}-1) \right),  
\end{eqnarray}
(Here we have attached a subscript $s$ to the structure function to 
stress that it is defined in the single cascade scheme whose meaning 
will become clear in the next section.) When $Q_0^2 \ll Q^2 $, we can
modify
\begin{eqnarray}
 D_s(n,Q^2)
 &=&\exp\left( {\alpha\over 2 \pi} \int_{\epsilon_Q}^{1} {d t\over t}
             \int^1_{\epsilon_K} d z P(1-z)[(1-z)^{n-1}-1) ]\right)
                 \nonumber \\
 &=&\exp\left( {\alpha\over 2 \pi} \int^1_{\epsilon_Q} 
     d z   \log({z\over \epsilon_Q}) P(1-z)[(1-z)^{n-1}-1) ]\right),
  \label{eq:mom4}
\end{eqnarray}
with $\epsilon_Q =Q^2_0/Q^2 $. From the last expression we find
 \begin{eqnarray}
      D_s(n,Q^2)&=&D(n,Q^2) D_{sf}(n) + O(\epsilon_Q),\nonumber\\
      D_{sf}(n)& =& \exp\left( {\alpha \over 2\pi} 
 \int_0^{1} d x P(x)\log(1-x)(x^{n-1}-1) \right),
  \label{eq:fnt}
\end{eqnarray}
where $D(n,Q^2)$ is given by Eq.(\ref{eq:mom3}) with $\epsilon=0$ and
$D_{sf}(n)$ is the finite contribution we have looked for.
In $x$-space, the equation (\ref{eq:fnt}) becomes a convolution 
integral with respect to $x$ as
\begin{equation}
      D_s(x,Q^2) =D(x,Q^2)\otimes  D_{sf}(x).   \label{eq:fnt2}
\end{equation}

In figure 1 one can see that $D_s(n,Q^2)$ is completely reproduced by
the Monte Carlo simulation when $Q_s^2=Q_0^2 $. If $Q_0^2\ll Q_s^2\ll 
Q^2$, however, the finite contributions vanish. This implies that 
$D(n,Q^2)$ should be coincides with the simulation(points with circle).
This is also demonstrated in the figure.


\vskip 0.5in
\noindent
{\bf  Section 3. The double cascade scheme}

The algorithm discussed in the previous section is for the case of the
single cascade scheme in $e^+ e^- $ annihilation. The characteristic 
of this scheme is that the vector to define the axial gauge is set 
parallel to the positron and as a result the shower develops only on 
the electron while the positron stays inactive.

In the program for the actual generator, however, we adopt the double 
cascade scheme\cite{double}. This is another formulation which is more 
suitable than the single cascade in order to assure the symmetric 
treatment of the radiation from $e^+$ and $e^-$ as both of these are
able to develop showers. In this section we discuss this scheme in some
details and look at what finite contribution comes out.

First we give a  review on the double cascade scheme,
where photons are radiated parallel to its parent electron or positron
but {\sl not anti-parallel} to it. To make the argument clear,
we take the simple case that only a single soft photon is
emitted. The momentum of the photon is denoted as $k$ and those of two 
(on-shell)electrons are $P_1$ and $P_2$, respectively.
 For simplicity we neglect the electron mass, 
that is $P_1^2=P_2^2=0$.
 The soft photon 
contribution $I$ is then given by
 \begin{eqnarray}
I ={e^2\over (2\pi)^3} \int {d^3k\over 2k_0}
 { 2(P_1 P_2)\over 2(P_1k)\  2(P_2k)}, 
 \end{eqnarray}
where $(ab)=a_0b_0-\vec{a}\cdot\vec{b}$. We use the light-cone 
momentum defined as
 \begin{eqnarray}
    P_{\pm} =(E \pm P_z)/\sqrt{2}, 
 \end{eqnarray}
and choose a frame in which we have
 \begin{eqnarray}
   P_{1\mu} =(P_{1+},0_-,\vec{0}_T),\ \ \ 
P_{2\mu} =(0_+,P_{2-},\vec{0}_T)  , \ \ \ 2(P_1P_2) = Q^2.
 \end{eqnarray}
Now two new variables $z$ and $t$, the fractions of the light-cone 
momentum of electrons, are introduced by
 \begin{eqnarray}
  z={(kP_2)\over(P_1P_2)}={ k_+\over P_{1+}},\ \ \ 
t= {(kP_1)\over(P_1P_2)}={k_-\over P_{2-}}. 
 \end{eqnarray}
In terms of these variables the soft photon contribution $I$ is 
rewritten as
 \begin{eqnarray}
 I = {\alpha \over 4\pi} \int_0^1 {dz \over z} 
 \int_0^1 {dt \over t}.  
 \end{eqnarray}
This integral is divergent so that a cutoff is needed. For this we
impose the lower limit to the transverse momentum squared of the photon,
$Q_0^2$.
\begin{eqnarray}
\vec{k}_T^2 &=& z 2(kP_1)= zt Q^2 \ \ > Q_0^2, \nonumber \\
I(Q^2_0) &=& {\alpha \over 4\pi} \int_0^1 {dz \over z} 
 \int_0^1 {dt \over t}\theta(zt-Q^2_0/Q^2).  
 \label{eq:soft1}
\end{eqnarray}
The events can be generated using these $z,t$ in the full phase space.
This generation method is the single cascade scheme.

In the C.M. frame we have $P_{1+}= P_{2-}=\sqrt{Q^2/2}$ so that
the $z$-component of the photon momentum is given by
 \begin{eqnarray}
        k_z={k_+ -k_-\over \sqrt{2}} = {\sqrt{Q^2} \over 2}(z-t).
 \end{eqnarray}
This shows that the photon with $ z > t $ is emitted parallel to 
the electron, while that with $ t > z$ parallel to the positron. 
Consequently it would be understood
that we can select the parent of a shower by looking at which 
of $z$ or $t$ is greater than the other. Thus a shower can develop from
any electron {\sl independently} and emitted photons are parallel 
to the electron or positron from which it has branched if one
imposes the restriction $ z > t $.
This is the double cascade scheme.
 
 The parton showers are based on the renormalization group equations
by which some physical quantity is calculated. It depends on the
process considered and for the case of the deep inelastic scattering, 
it is the distribution on the Bjorken variable. The parton showers must 
give the same distribution which can be calculated in the
analytic approach.
  
In the single cascade scheme the fraction of the light-cone variables
is equal to the Bjorken variable. In the double cascade scheme this 
is not valid and we must find what combination of which variables 
generated in parton showers is equal to the Bjorken variable.

Using the example of the single soft photon emission, we will look
 at the problem in details.
Then the variable $x_{Bj}$ can be introduced
by the following way:
 \begin{eqnarray}
 P_1 + q = P_2 + k, \ \ \  2(P_1 q) + q^2 = 2(P_2 k), 
 \end{eqnarray}
 \begin{eqnarray}
 {1\over x_{Bj}}= {2(P_1 q)\over -q^2} = 1+{(P_2 k)\over -q^2}.
 \end{eqnarray}
Since $-q^2 = 2(P_1P_2)(1+t-z) $, we find
 \begin{eqnarray}
x_{Bj} \equiv 1- z/(1+ t)= 1- z +O(zt).
 \end{eqnarray}
For small $t$ the Bjorken variable is then determined by only $z$.
This kinematics corresponds to the single cascade scheme,
\begin{eqnarray}
 D(x_{Bj})&=&\delta(x_{Bj}-1)( 1-I(Q^2_0)) \nonumber  \\
    &&+{\alpha \over 4\pi} \int_0^1 {dz \over z}
 \int_0^1 {dt \over t}\theta(zt- Q^2_0/Q^2)
 \delta(x_{Bj}-1+z). \nonumber \\
 \label{eq:soft3}
\end{eqnarray}

In the double cascade scheme, on the other hand, the inequality between
$z$ and $t$ directly connected to the choice that either of which 
electrons emits the photon. In other words, when $ z < t $ or 
$(P_2k)<(P_1 k)$, the role of $z$ and $t$ should be interchanged. 
Hence in our example of the single soft photon emission, 
the distribution over $x_{Bj}$ becomes
\begin{eqnarray}
 D(x_{Bj})&=&\delta(x_{Bj}-1)( 1-I(Q^2_0)) \nonumber  \\
    &&+{\alpha \over 4\pi} \int_0^1 {dz_1 \over z_1} 
 \int_0^1 {dt_1 \over t_1}\theta(z_1t_1- Q^2_0/Q^2)
 \theta(z_1-t_1)\delta(x_{Bj}-1+z_1) \nonumber \\
    &&+{\alpha \over 4\pi} \int_0^1 {dz_2 \over z_2} 
 \int_0^1 {dt_2 \over t_2}\theta(z_2t_2- Q^2_0/Q^2)
 \theta(z_2-t_2)\delta(x_{Bj}-1+t_2).
 \label{eq:soft2}
\end{eqnarray}
The first term would be easily understood if one notices that
the sum of contributions from real photon emissions and loop 
diagrams is finite(no infrared and no collinear singularity). 
In this expression the subscript 1(2) designates the 
electron with momentum $P_1(P_2)$ and each contribution corresponds to
the radiation from them. It is clear
then by superposing two showers defined with the constraint $z>t$, we 
get all the contributions in the full phase space. Also the radiated 
photons are parallel to the parent electrons. It is this constraint 
that allows us to draw a picture of jets, a cluster 
of electrons and photons flowing in the same direction.
One should notice that in Eq.(\ref{eq:soft2}), $x_{Bj} $ can be replaced  
by $ (1-z_1)(1-t_2) $.

Let us apply the above argument to showers with indefinite number of 
photons. Here we will repeat the main points to be argued.
In the double cascade scheme, parton showers develop with
 {\sl constraint}
$ (1-x) \geq t$, where $x$ is the fraction for the electron, not for 
photons. Then what is the physical quantity that can be calculated by 
the renormalization group equation and what is the combination of the
variables that equal to this physical quantity?

For the deep inelastic scattering it is obvious that the answer for 
the first question is the Bjorken variable. Then we study the second 
question. For this we specify the process as
 \begin{eqnarray}
 e(P_1) +\gamma(q) \rightarrow  e(p_1)+X +\gamma(q)\rightarrow 
e'(p_2) +X \rightarrow X'.  
 \end{eqnarray}
Here $X$ denotes anything. A virtual electron with $p_1$ collides with 
the virtual photon with $q$ and turns into another virtual electron 
with $p_2$. The momentum conservation gives
 \begin{eqnarray}
  p_1^2+2(p_1q)+ q^2 = p_2^2  .  
 \end{eqnarray}
In the frame
 \begin{eqnarray}
 P_{1-}=  0, \ \ q_+ = - p_{1+}, \ \ \vec{q}_T = \vec{0},\ \ \  
x_b={p_{1+}\over P_{1+}}, 
 \end{eqnarray}
we have
 \begin{eqnarray}
 x_{Bj}\equiv{-q^2\over2(P_1q)} = 
x_b-{p_2^2\over 2(P_1q)} -{\vec{p}_1^2\over 2(P_1 q)},
 \end{eqnarray}
by noting 
 \begin{eqnarray}
 { 2(p_1 q) \over 2(P_1 q)} = x_b -{p_1^2 + \vec{p_1}^2 \over 2(P_1q)}.
 \end{eqnarray}
Since $\vert\vec{p}_1\vert = O(1-x_b)p_1^2 $, it can be neglected
as well as the product of $ x_b$ and $p_2^2/(2(P_1q))$.
We find
 \begin{eqnarray}
 x_{Bj}\simeq x_b\left(1-{p_2^2\over 2(P_1q)}\right)=x_b(1-t_2).
 \end{eqnarray}
In the double cascade scheme $p_2^2 >0 $. This means that the electron
$p_2$ has a potential to emit photons and we must include contributions
from showers of the scattered electron. 

Next we have to prove that the distribution over $x_b(1-t_2)$ in fact
agree with the solution of the renormalization group equation.
In the course of the proof one will see that this combination of 
variables is suitable for our discussion. Also after the proof, one will
find the finite contribution we are looking for.

The $x_b$-distribution is nothing but the distribution in the single 
cascade but with the constraint $(1-x) \geq t$ inside of the integral
in exponential. 
\begin{equation}
   D_x(n,Q^2)=\exp\left( {\alpha \over 2 \pi } \int^{Q^2}_{Q_s^2}
    { d K^2 \over K^2 } \int_0^{1-\epsilon_K } P(x)( x^{n-1} - 1)
    \theta(1-x - t) \right) . 
\end{equation}
The expansion of the distribution over $t_2$ is
 \begin{eqnarray}
 &&D_t(t_2,Q^2) = \Pi_c(Q^2,Q^2_0)\delta(t_2)  \nonumber \\
&&+{\alpha \over 2\pi} \int_0^1 { dt \over t} \int_0^1 dx P(x)
 \theta(1-x-t)\theta((1-x)t-Q^2_0/Q^2) \Pi_c(Q^2,tQ^2)
 \delta(t-t_2) . \nonumber \\
&&
 \end{eqnarray}
\noindent
Here $ \Pi_c(Q^2,K^2)$ is the probability of non emission with
the constraint, i.e. the probability in the double cascade scheme.
\begin{equation}
 \Pi_c(Q^2,K^2)=
 \exp\left( -   {\alpha \over 2\pi} \int_{K^2/Q^2}^1 { dt \over t}
 \int_0^1 dx P(x)
 \theta(1-x-t)\theta((1-x)t-Q^2_0/Q^2) \right).
\end{equation}
The moment with respect to $t_2$ is defined by  
\begin{equation}
D_t(n,Q^2)=\int_0^1 dt_2 (1-t_2)^{n-1} D_t(t_2,Q^2).
\end{equation}
Noting that $(1-t)^{n-1}\simeq\theta(1/n-t)$ for large $n$, 
we find 
 \begin{eqnarray}
     D_t(n,Q^2)&=&\exp\left( {\alpha \over 2 \pi } \int^{Q^2}_{Q_s^2}
    { d K^2 \over K^2 } \int_0^{1-\epsilon_K }dx P(x)( (1-t)^{n-1} - 1)
    \theta(1-x - t) \right) , \nonumber \\
 &&\qquad\qquad\qquad\epsilon_Q= Q_0^2/Q^2, \ \  t=  K^2/Q^2 .
          \label{eq:dbl2x}
\end{eqnarray}
\noindent
We have obtained closed forms for the moment distributions using some 
approximation. Since we know that the product of $x_b$ and $1-t_2$ is 
the Bjorken variable, the moment of $x_{Bj}$, $D_d(n,Q^2)$ is the 
product of the moment $D_x(n,Q^2)$ with respect to $x_b$ and 
$ D_t(n,Q^2)$ to $t_2$,
\begin{equation}
   D_d(n,Q^2)=D_x(n,Q^2) D_t(n,Q^2). \label{eq:dbl1}
\end{equation}

Next we compare these moment distributions with those obtained by
Monte Carlo simulation, which will confirm our analysis.
First the distribution of $x_b$ will be discussed.
The analytic form of the moment distribution can be
calculated easily.
\begin{eqnarray}
   D_x(n,Q^2)&=&\exp\left( {\alpha \over 2 \pi } \int^{Q^2}_{Q_s^2}
    { d K^2 \over K^2 } \int_0^{1-\epsilon }dx P(x)( x^{n-1} - 1)
    \theta(1-x - t) \right)  \nonumber \\
 &=&\exp\left( {\alpha \over 2 \pi } \int_{\sqrt{\epsilon_Q}}^1dz
\log({z^2 \over\epsilon_Q} )
P(1-z)[ (1-z)^{n-1} - 1]\right).
             \label{eq:dbl3}
\end{eqnarray}
Figure 2 shows the Monte Carlo results for $x_b$-distribution in the 
double cascade scheme and those obtained from the analytic form
Eq.(\ref{eq:dbl3}). The agreement justifies our formulation.

Now let us calculate the finite contributions in the double cascade 
scheme. First we notice
\begin{eqnarray}
 {1\over t}P(1-z)={1\over z}P(1-t)+{2-t\over z}
            +{-2+z\over  t} .  
\end{eqnarray}
The moments in the scheme are then
 \begin{eqnarray}
  D_d(n,Q^2)&=& D_x(n,Q^2)  D_t(n,Q^2)  \nonumber \\
&=&\exp\left( {\alpha \over 2 \pi } \int^{1}_{\epsilon_Q }
    { d t \over t } \int_0^{1-\epsilon_Q/t } dx P(x)( x^{n-1} - 1)
                                 \right)
     \nonumber \\
&&\times  \exp\biggl( {\alpha \over 2 \pi } \int_0^1dt \int^1_0 dz
\left({2-t\over z}-{2-z\over t }\right)\theta(zt -\epsilon_Q)
     \theta(z - t) \nonumber  \\
&&\qquad\qquad\qquad \times   [(1-t)^{n-1}-1]\biggr) .
             \label{eq:dlb4}
\end{eqnarray}
The first exponential on the right-hand side is the same
as the moments $ D_{s}(n,Q^2)$ appearing in the single cascade scheme. 
In the latter the integration over $z$ is done and O($\epsilon_Q$)
terms are neglected. After changing variable $t$ by $z$, we get
 \begin{eqnarray}
&&D_{d}(n,Q^2)=  D_{s}(n,Q^2) \nonumber \\
&&  \times\exp\left({\alpha \over 2 \pi } \int_0^1dz
\left((-2+z)\log(z) -{3\over 2z}+2-{z\over 2}\right)
                              [ (1-z)^{n-1} - 1]\right).
          \label{eq:dd2}
\end{eqnarray}
Hence we finally find
 \begin{equation}
   D_{d}(n,Q^2)= D(n,Q^2)D_{df}(n),
          \label{eq:dd}
\end{equation}
where the finite term is given by
 \begin{eqnarray}
D_{df}(n)&=& D_{sf}(n)
  \exp\left({\alpha \over 2 \pi } \int_0^1dz
\left((-2+z)\log(z) -{3\over 2z}+2-{z\over 2}\right)
                        [ (1-z)^{n-1} - 1]\right),
\nonumber  \\ 
 & =& \exp\biggl( {\alpha \over 2 \pi } \int_0^1 dz
[ \log(z)P(1-z) + 
((-2+z)\log(z) -{3\over 2z}+2-{z\over 2})]    \nonumber  \\   
 &&\qquad\qquad\times\lbrack (1-z)^{n-1} - 1 \rbrack \biggr). 
          \label{eq:df}
\end{eqnarray}

This analytic form can be compared with the Monte Carlo results in
Fig.2. The agreement confirms definitely our discussions. Hence
to get the solution of the renormalization group equation, $ D(n,Q^2)$,
we have to subtract the finite contribution in Eq.(\ref{eq:df})
or calculate the ratio of distributions with different $Q^2$.
Thus we have completed the analysis of the deep-inelastic scattering.

In $e^+ e^- $ annihilation, a more complicated situation arises due
to its kinematics. It is the fact that the squared mass of the virtual 
photon, $q^2$, is calculated by 
momentum fractions $x_1, x_2 $ of the electron and the positron.
\begin{eqnarray}
   q^2 = x_1 x_2 s ,  
\end{eqnarray}
\noindent
where $s$ is the square of the energy in the center-of-mass system.

In the double cascade scheme we consider the following process allowing
both electron and positron to radiate photons,
\begin{eqnarray}
 e^-(P_1)+e^+(P_2) \rightarrow
e^-(p_1)+e^+(p_2)+ X  \rightarrow \gamma(q)+ X.
\end{eqnarray}
\noindent
By this equation we mean that the spacelike virtual electron($p_1$) and
positron($p_2$) annihilates into the virtual photon($q$).
In the frame that $P_{1\mu}=(P_{1+},0_-,\vec{0}_T) $
$P_{2\mu}=(0_+,P_{2-},\vec{0}_T)$ a simple algebra gives
\begin{eqnarray}
 q^2 &=&(p_1+p_2)^2 \nonumber \\
     &=&p_1^2+p_2^2 +2x_{b1}x_{b2}(P_{1+}P_{2-})
+{(p_1^2+\vec{p_1}^2) (p_2^2+\vec{p_2}^2) \over 2x_{b1}x_{b2}
(P_{1+}P_{2-})} -2\vec{p_1}\cdot\vec{p_2},
\end{eqnarray}
where
\begin{eqnarray}
  x_{b1} ={ p_{1+} \over P_{1+}}, \ \ \
  x_{b2} ={ p_{2-} \over P_{2-}} . 
\end{eqnarray}
If one neglects the terms of order $(1-x_b)t$, one finds
\begin{eqnarray}
  q^2 = x_{b1} x_{b2}(1-t_1)(1-t_2) s  , 
\end{eqnarray}
where $t_1 = \mid p_1^2 \mid /s, t_2= \mid p_2^2 \mid /s $.

In the deep-inelastic scattering one calculates
the Bjorken variables 
in terms of the light-cone fraction of the electron in
one shower and the squared virtual mass
of the electron in another independent shower.
$x_{Bj} = x_{b1}(1-t_2) $.
 While in the annihilation process both of 
the light-cone fraction and the squared virtual mass determined by the 
generated shower are needed. Hence the moments with respect to the 
variable $x_b(1-t)$ are required for one shower. These are given by
\begin{eqnarray}
&&\Pi(Q^2,Q^2_0)+ \int_0^1 {dt \over t} (1-t)^{n-1} \Pi(Q^2,tQ^2)
                       \nonumber\\
&&\qquad\qquad\times{\alpha \over 2 \pi }
\int_0^1 dx P(x)x^{n-1}\theta(1-x - t)\theta((1-x)t-Q^2_0/Q^2) 
                                                \nonumber \\
&&\times
  \exp\left( {\alpha \over 2 \pi } \int^{tQ^2}_{Q_0^2}
    { d K'^2 \over K'^2 } \int dy P(y)( y^{n-1} - 1)
\theta(1-x - K'^2/Q^2)\theta((1-x)K'^2-Q^2_0) 
\right).  \nonumber \\
&&
 \end{eqnarray}
By using $ (1-t)^n \sim \theta(1/n -t) $, it turns out that 
the moments of $ Q^2 /s $ is given by the square of $D_a(n,Q^2)$.
 \begin{eqnarray}
D_a(n,Q^2)&=& \int^1_0 d \xi \xi^{n-1} D_a(\xi, Q^2) \nonumber \\
&&\times  \exp\left( {\alpha \over 2 \pi } \int^{Q^2}_{Q_0^2}
{ d K^2 \over K^2 } \int_0^{1-\epsilon_Q } dx P(x)( (x(1-t))^{n-1} - 1)
    \theta(1-x - t) \right),   \nonumber \\
   && \xi=Q^2/s, 
\label{eq:annh}
\end{eqnarray}
\noindent
The finite contributions $D_{af}(n)$ is then given by
 \begin{eqnarray}
   D_{a}(n,Q^2)&=& D(n,Q^2) D_{af}(n,Q^2) \nonumber\\
D_{af}(n,Q^2)&=& \exp\biggl( {\alpha \over 2 \pi } 
    \int_0^1 dx \lbrack P(x) \log(1-\sqrt{x}) 
 -2{1-x\over x} \log{\sqrt{1-x}\over 1+\sqrt{1-x}} \nonumber\\
    &&\qquad\qquad+{1-x\over 2}\log(1-x) \rbrack (x^{n-1}-1)  \biggr),
          \label{eq:annh2}
\end{eqnarray}
for $\epsilon_Q \ll 1 $.

\noindent
\vskip 0.5in
\noindent
{\bf  Section 4. The model}

Discussions in the previous sections suggest that it is easy
to correct the finite contributions in the moment space.
For the annihilation process
\begin{eqnarray}
  D(n,Q^2) = D_a(n,Q^2)/D_{af}(n) .
\end{eqnarray}
\noindent
Since $ D_{af}(n)^{-1} < 1 $, its inverse transform 
$\overline{D}_{af}(x)$ is well defined, 
\begin{eqnarray}
 \overline{D}_{af}(x) ={1\over 2\pi i}
\int_{-i \infty +c}^{i \infty +c} x^{-n} {1\over D_{af}(n)}.
\end{eqnarray}
\noindent
We will show how to take into account  these corrections
 in $e^+ e^- $ annihilation process.
As a simple analytic form is not easy to get for $\overline{D}_{af}(x)$,
we make the following approximation
\begin{equation}
\overline{D}_{af}(x)  \simeq
\overline{D}_{af}^A(x)  \equiv C\delta(1-x) + \theta(x_f -x)F(x),
\end{equation}
where the function $F(x)$ is given by
\begin{eqnarray}
 F(x)&=&{ \alpha \over 2 \pi } 
    \biggl\lbrack P(x) \ln(1-\sqrt{x}) 
 -2{1-x\over x} \ln{\sqrt{1-x}\over 1+\sqrt{1-x}} \nonumber \\
    &&\qquad\qquad+{1-x\over 2}\ln(1-x) \biggr\rbrack 
\exp\left(-{\alpha \over 2 \pi }\ln^2(1-x) \right).
          \label{eq:app}
\end{eqnarray}
\noindent
Here $ 1-x_f \ll 1$ is assumed and the constant $C$ is fixed by 
the requirement that
\begin{eqnarray}
  \int_0^1 dx \overline{D}_{af}^A(x) = 1. 
\end{eqnarray}
The parameter $x_f$ is fixed by the condition that a photon with energy
fraction $1-x_f$ cannot be observed in the detectors.
In usual experiments the limit of the measurable energy is in the order
of 10 MeV. From this fact
in the program we set $x_f=1 -10^{-4} $.
If $1-x_f$ is extremely small, the approximation of 
$ \overline{D}_{af}^A(x) $ for $\overline{D}_{af}(x)$
is not justified.
Then we add the following algorithm before developing showers.

 At $ Q_s^2 $, we choose $x$ according to $\overline{D}_{af}^A(x) $.
 If $x \not=  1 $, a photon with the momentum fraction $1-x $
 is emitted parallel to the electron and the energy fraction of the
 electron is $x $. In the generator $ Q_s^2 \gg Q^2_0 $ is assumed.

Next we will present results by the improved QEDPS and compare
them with the analytic results. Since the coupling constant $\alpha$ 
is small, we make a comparison on the integrated radiator over small 
$1-x$, that is,
\begin{eqnarray}
 R_i(x,Q^2) &=& \int_x^1 dy R(y,Q^2),\\
 R(x,Q^2) &=& \int_0^1 dx_1 dx_2 D(x_1,Q^2)D(x_2,Q^2) \delta(x-x_1x_2).
\end{eqnarray}
In the Monte Carlo we count the fraction of events with 
$ s_{effective} \geq x s $. In the simulation we assumed the following 
values for the parameters.
\begin{eqnarray}
 \alpha =1/137, Q_s^2=10^{-6} \mbox{GeV}^2, Q^2=10^4 \mbox{GeV}^2, 
  Q_0^2 =10^{-8}\mbox{\GeV}^2.   
\end{eqnarray}
The results are shown in figure 3. 

In analyzing the experimental data people use very sophisticated 
radiators, which involve the higher order corrections and others.
A review on these will be found in ref.\cite{rad}. We call the radiator
given there $R_{Suppl}(x,Q^2)$. This and $R(x,Q^2)$ are plotted in 
figure 3, in which one still finds a small discrepancy on the 
order of $1 \%$. We should note, however, that
\begin{eqnarray}
   \int_0^1 dx R(x,Q^2 ) = 1, 
\end{eqnarray}
while
\begin{eqnarray}
   \int_0^1 dx R_{Suppl}(x,Q^2 ) = 
1+ {\alpha \over \pi}\left({\pi^2\over 3}-{1\over 2}\right)=C_a. 
\end{eqnarray}
We introduce a modified $R_M(x,Q^2)$ as
\begin{eqnarray}
  R_M(x,Q^2) = R(x,Q^2) C_a. 
\end{eqnarray}
Then we plot $R_M(x,Q^2)$ and $R_{Suppl}(x,Q^2)C_a$ to
find a satisfactory agreement whose accuracy is less than $0.1 \%$.
Hence the normalization is fixed by  $C_a $ instead of unity in 
the shower.

\vskip 0.5in
{\bf Section 5.   Summary and discussions}

In this paper we present the improved version of QEDPS, a generator
for radiative corrections to the processes in $e^+ e^- $ annihilation.
This new model has been obtained from the old one, that has been 
proposed by the authors a few years ago, by correcting the finite 
contributions originated from our shower algorithm. It was shown that
these contributions appear when the infra-red cutoff is $K^2$-dependent.

In the present work we did not take account of the running effect for
the QED coupling $\alpha$. If one applies this study to improve the 
finite contributions in QCD showers, the running effects should be 
considered. We would like to discuss this point in a separated paper.

\vskip 0.5in
{\bf Acknowledgements}

We would like to thank  our colleagues of KEK working 
group(Minami-Tateya) and in LAPP for their interest and discussions. 
This work has been done under the collaboration between KEK and LAPP 
supported by Monbusho, Japan(No. 07044097) and CNRS/IN2P3, France.
\noindent

\eject

\eject

{\bf Figure Captions}

Fig.1~~  Moments of the structure functions.
The solid curve is the correct results obtained by the renormalization
group equation. The dashed curve is given by the analytic form of
$D_s(n,Q^2)$. Diamond(circle) data points are Monte Carlo results in 
the single cascade scheme for $Q_s^2=Q_0^2( Q^2_s \gg Q_0^2 )$.
Parameters are $ Q^2=10^4 \GeV^2, Q^2_s=10^{-6}GeV^2, \alpha=1/20$.
For the Monte Carlo data of $ Q^2_s \gg Q_0^2 $, we choose
$Q^2_0=10^{-8}\GeV^2$.

 Fig.2~~  Moments of the structure functions in the double
 cascade scheme.
The solid(dashed) curve is calculated analytically by $D_x(n,Q^2)$
 ($D_x(n,Q^2)D_t(n,Q^2)$).
 Diamond(circle) data points of the Monte Carlo simulation 
corresponds to $x_b(x_{Bj}=x_b(1-t))$.
The long-dashed curve presents the results of $D(n,Q^2)$.
 Parameters are the same as Fig.1.

 Fig.3~~  Differences between the Monte Carlo data and
       analytic expression for the integrated radiator,
       $ \int^1_x dy R(y,Q^2) $.   
       Parameters are the same as Fig.2 except $\alpha $.
       Here $\alpha= 1/137 $. One million events are generated.

\begin{figure*}[htb]
\centerline{\epsfbox{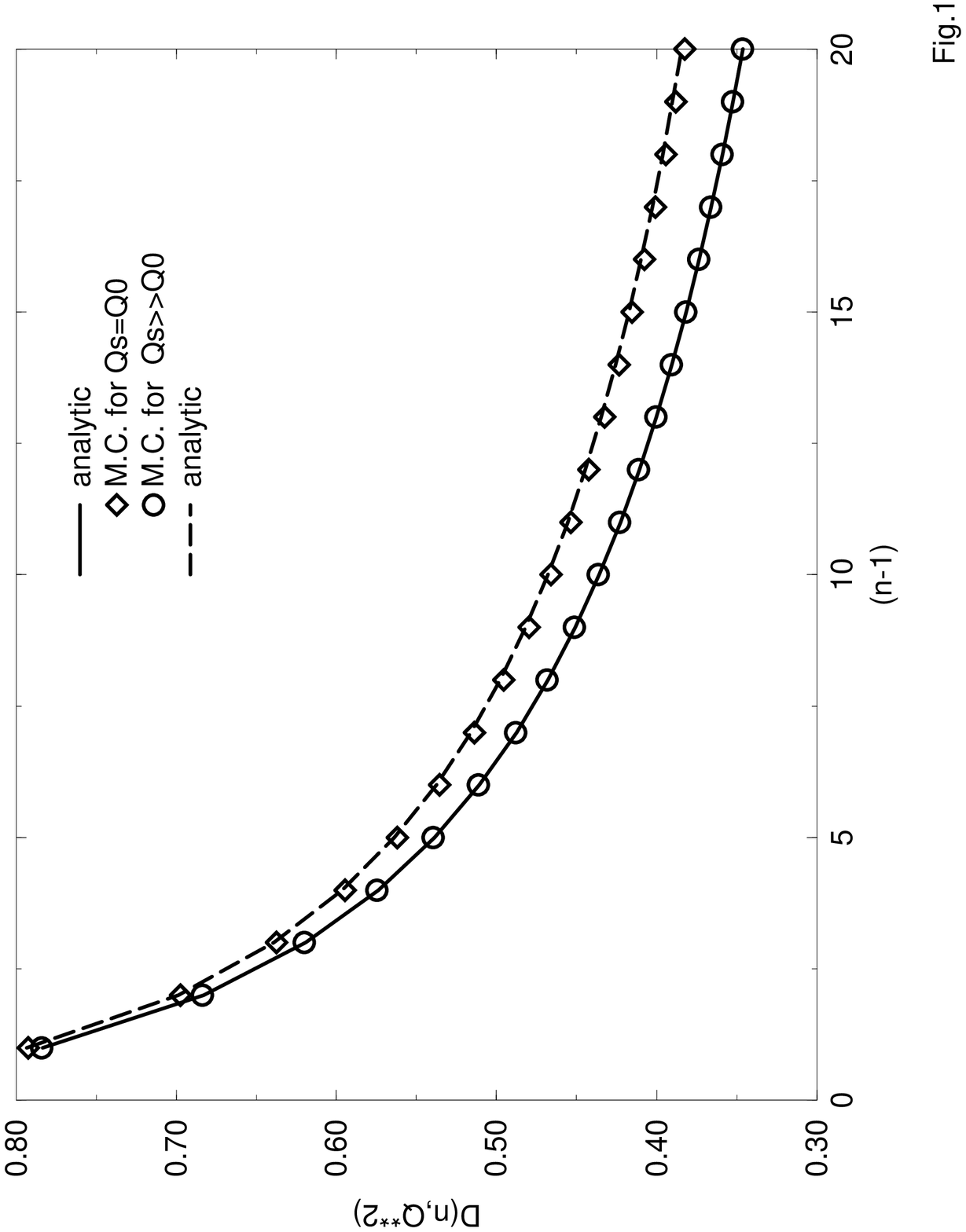}}
\end{figure*}
\begin{figure*}[htb]
\centerline{\epsfbox{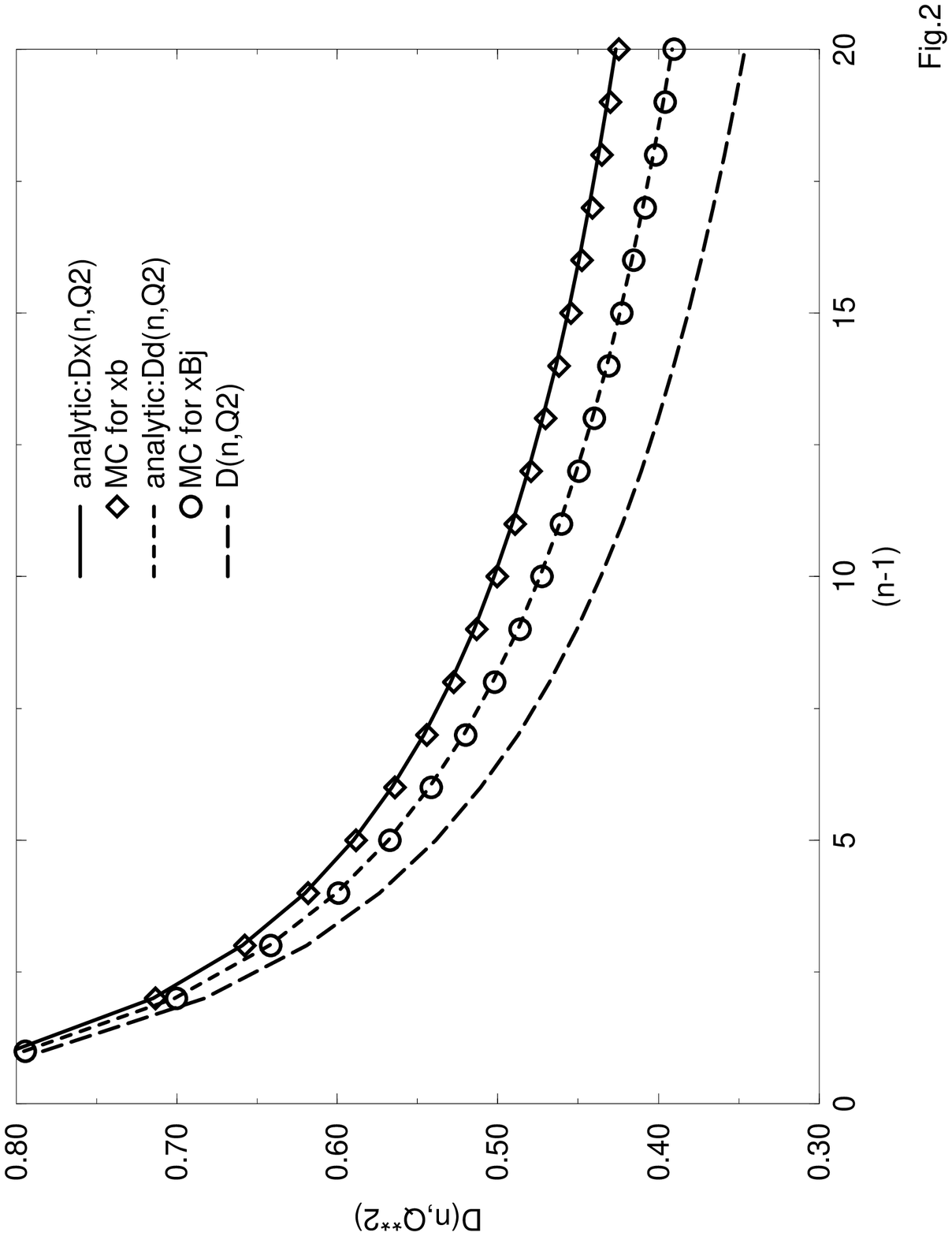}}
\end{figure*}
\begin{figure*}[htb]
\centerline{\epsfbox{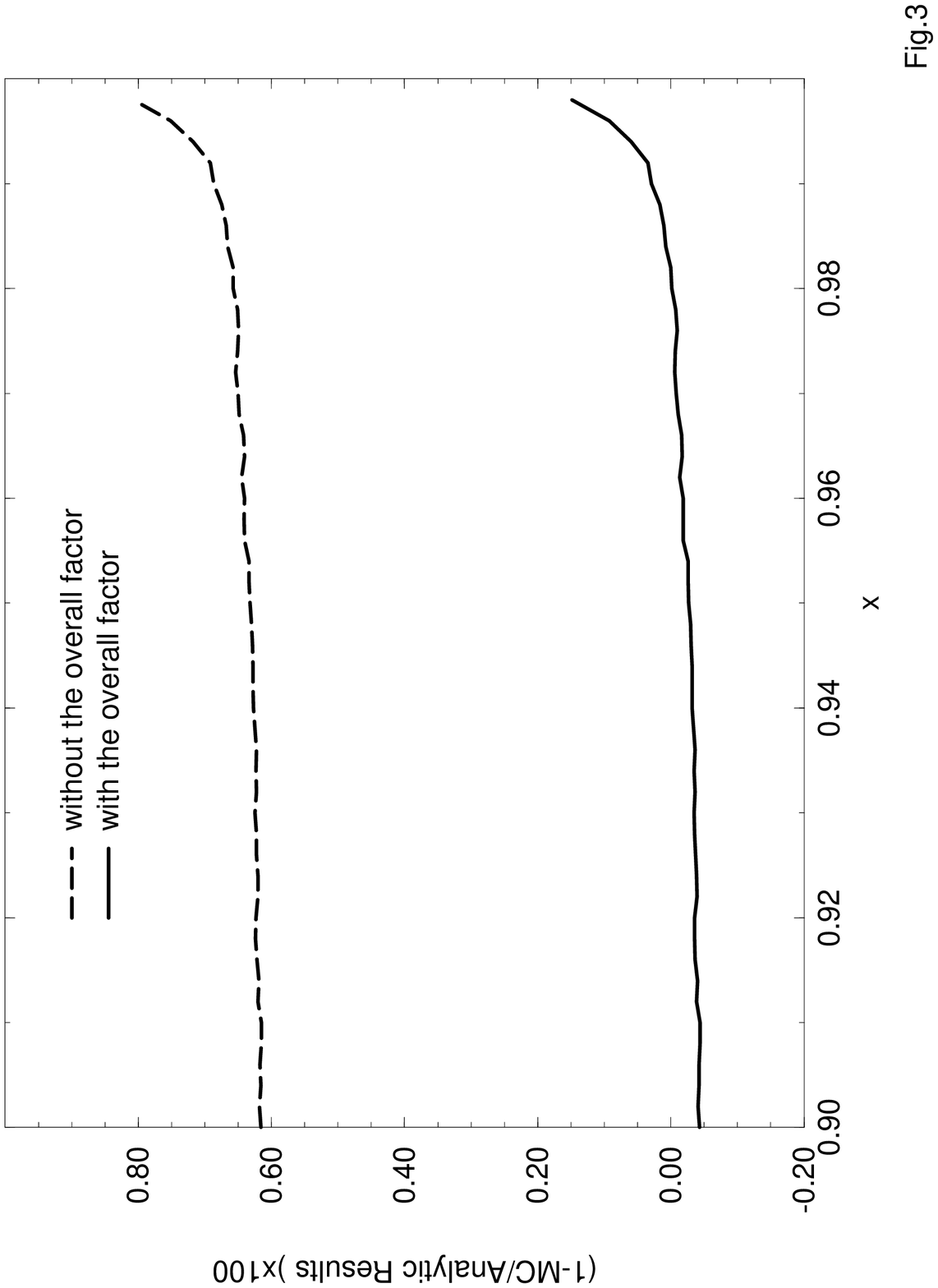}}
\end{figure*}
\end{document}